\documentclass[format=sigconf]{acmart}
\usepackage[utf8]{inputenc}
\usepackage{enumitem}
\usepackage{times}

\title{In-network Neural Networks}
\author{Giuseppe Siracusano, Roberto Bifulco\\ NEC Laboratories Europe}
\date{December 2017}

\setcopyright{none}
\renewcommand\footnotetextcopyrightpermission[1]{} % removes footnote with conference information in first column
\begin{document}

\maketitle

\section{Introduction}
Network devices, such as switches and routers, process data at rates of terabits per second, forwarding billions of network packets per second. 
Recently, such devices' switching chips have been enhanced to support new levels of programmability~\cite{forwardingMetamorphosis}.  Leveraging these new capabilities, a switching chip's packets classification and modification tasks can now be adapted to implement custom functions. For example, researchers have proposed approaches that rethink load balancers~\cite{silkroad}, key-value stores~\cite{NetCache}, and consensus protocols~\cite{SwitchPaxos} operations. In general, there is a trend to offload to the switching chips (parts of) functions typically implemented in commodity servers, thereby achieving new levels of performance and scalability. 

These solutions often offload some data classification tasks, encoding relevant information, e.g., the \textit{key} of a key-value store entry~\cite{switchKV}, in network packets' headers. Unlike packets' payload, the header values can be parsed and processed by the switching chips, which perform classification using lookup tables. 
% The tables are implemented both with SRAMs, to exactly match on header's values, and ternary content associative memories (TCAMs), to match on ranges of values. 
While providing very high throughput, lookup tables need to be filled with entries that enumerate the set of values used to classify packets, and therefore the table's size directly correlates to the ability to classify a large number of patterns. Unfortunately, the amount of memory used for the tables is hard to increase, since it is the main cost factor in a network device's switching chip~\cite{forwardingMetamorphosis}, accounting for more than half of the chip's silicon resources. 

In this paper, we explore the feasibility of using an artificial neural network (NN) model as classifier in a switching chip, as a complement to existing lookup tables. A NN can better fit the data at hand, potentially reducing the memory requirements at the cost of extra computation~\cite{learnedIndexes}. Here, our work builds on the observation that, while adding memory is expensive, adding circuitry to perform computation is much cheaper. For reference, in a programmable switching chip the entire set of computations is implemented using less then a tenth of the overall chip's area.

% Kraska et al.~\cite{TheCaseforLearnedIndexStructures} make the case for using neural network models to replace index structures, providing in several cases a more efficient memory usage than state of the art solutions. Accordingly, in this paper we argue that a switching chip's classification task could be replaced by a neural network model, in order to save on memories' size.

% Nonetheless, common wisdom suggests that the expensive computations required by a neural network would make them unsuitable in a context where billions classifications per second are required. In this preliminary work we show that it is in fact possible to implement simple neural network models for such task, even with unmodified today's network devices.

% Nonetheless, 
To this end, we implement N2Net, a system to run NNs on a switching chip. We provide the following contributions: first, we show that a modern switching chip is already provided with the primitives required to implement the forward pass of quantized models such as \textit{binary neural networks}, and that performing such computation is feasible at packets processing speeds; second, we provide an approach to efficiently leverage the device parallelism to implement such models; third, we provide a compiler that, given a NN model, automatically generates the switching chip's configuration that implements it. Our experience shows that current switching chips can run simple NN models, and that with little additions a chip design could easily support more complex models, thereby addressing a potentially larger set of applications.

\vspace{0.05in}
\noindent \textbf{Use cases} At the time of writing we are still in the process of implementing full-fledged applications, thus, we just mention our two initial use cases, postponing to a later publication a throughout technical description. First, similar to ~\cite{learnedIndexes}, we envision the use of a neural network classifier to implement packet classification inside the chip, e.g., to create large white/blacklist indexes for Denial of Service protection. Second, the outcome of the NN classification can be encoded in the packet header and used in an end-to-end system, to provide "hints" to a more complex processor located in a server, e.g., on how to handle the packet's payload to optimize data locality/cache coherency or to support load balancing~\cite{flexswitches}.

\vspace{0.05in}
\noindent\emph{To encourage the community in exploring more use cases, we are in the process of making N2Net code publicly available.}

\section{N2Net}

\begin{figure}
    \centering
    \includegraphics[width=1\linewidth]{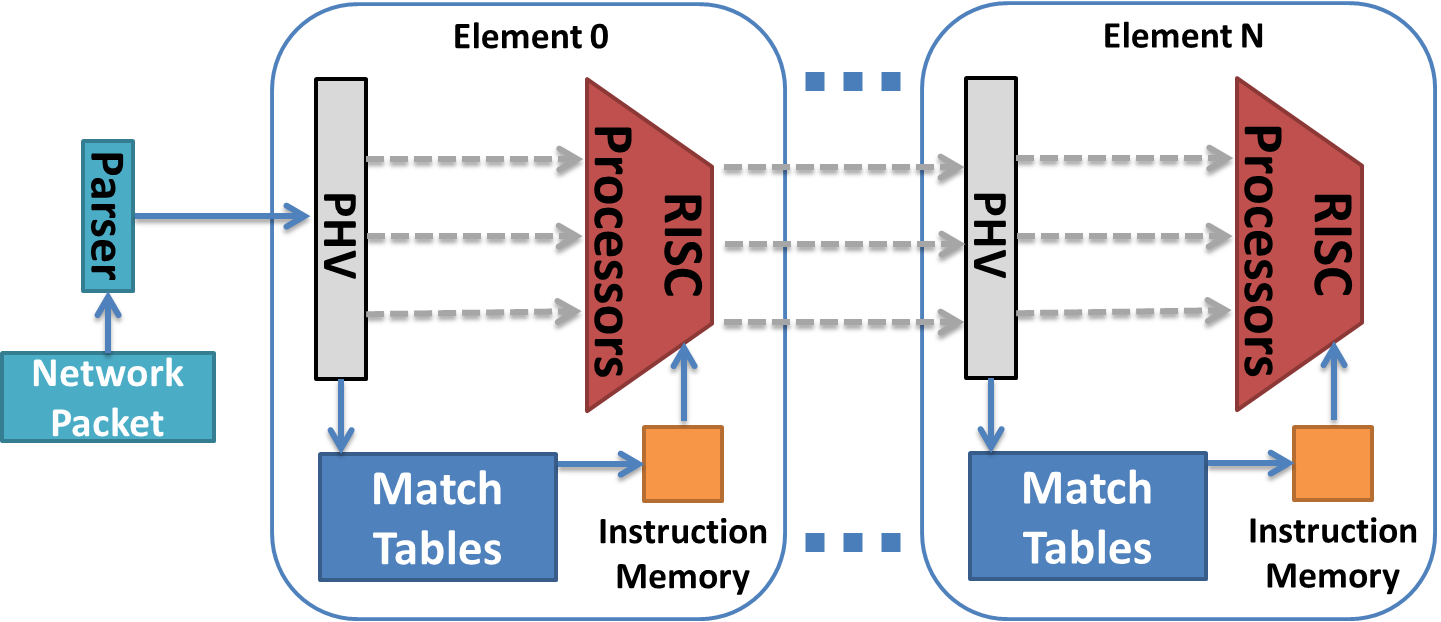}
    \vspace{-0.5cm}
    \caption{A schematic view of a switching chip's pipeline.}
    \vspace{-0.3cm}
\label{fig:switching_chip}
\vspace{-0.3cm}
\end{figure}

\noindent\textbf{Switching chip primer}
Switching chips process network packets to take a forwarding decision, e.g., forward to a given port or drop. They also perform transformations to the packet header values, e.g., to decrement a time-to-live field. We use RMT~\cite{forwardingMetamorphosis} as representative model of state-of-the-art switching chips (Cf. Fig.~\ref{fig:switching_chip}). When a packet is received, an RMT chip parses several 100s bytes of its header to extract protocol fields' values, e.g.,  IP addresses. These values are written to a packet header vector (PHV) that is then processed by a pipeline of \textit{elements} that implement match-action tables. Each element has a limited amount of memory to implement lookup tables (the match part), and hundreds of RISC processors that can read and modify the PHV in parallel (the action part). The values in the PHV are used to perform table lookups and retrieve the instruction the processors should apply. To provide very high performance, these processors implement only simple operations, such as bitwise logic, shifts and simple arithmetic (e.g., increment, sum). Using a language such as P4~\cite{p4}, the chip can be programmed to define the parser logic and the actions performed on the PHV. In particular, the actions are defined as a combination of the simpler primitives mentioned earlier. 

\vspace{0.1in}
\noindent\textbf{Design} The limited set of arithmetic functions supported by a switching chip does not enable the implementation of the multiplications and activation functions usually required by a NN. However, simplified models designed for application in resource-limited embedded devices, such as \textit{binary neural networks} (BNNs), do not require such complex arithmetic, especially during the forward pass~\cite{trainingBNN}. In our case, we select models that only use bitwise logic functions, such as XNOR, the Hamming weight computation (POPCNT), and the SIGN function as activation function. While research in these models is at its early stages, it shows already promising results~\cite{bnn, XNOR-net, bitwiseNN}.

\begin{figure} [t]
    \centering
    \includegraphics[width=1\linewidth]{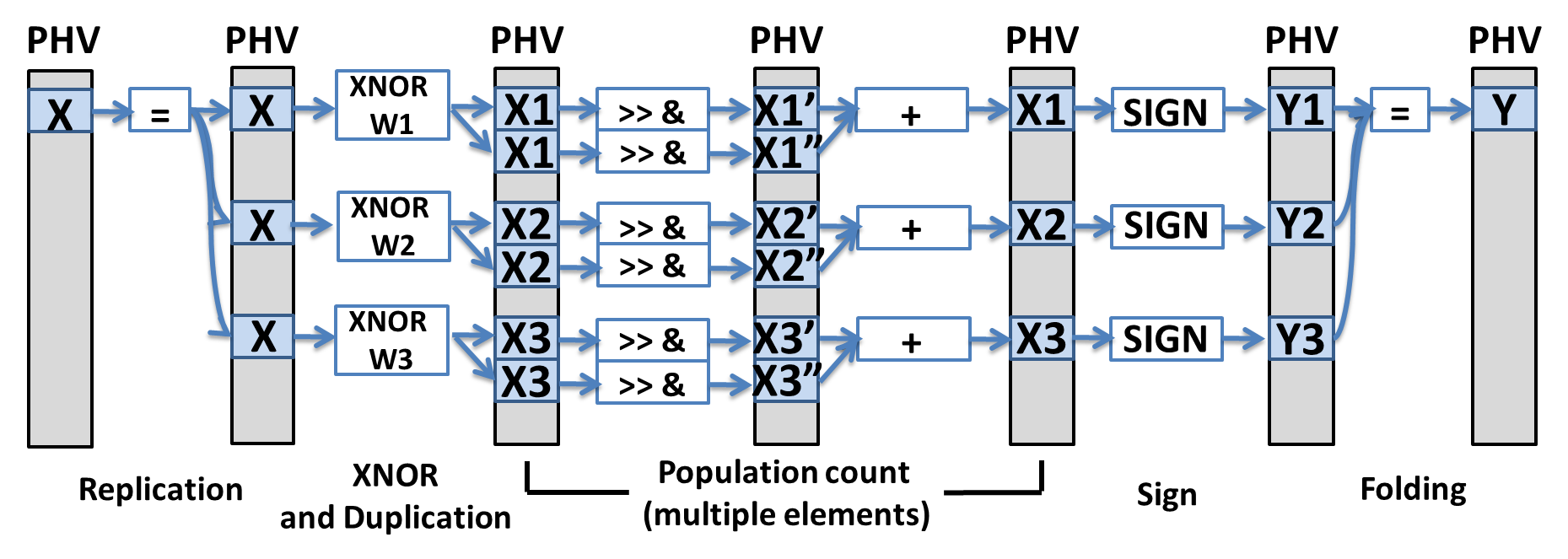}
    \vspace{-0.6cm}
    \caption{Implementation of a 3 neurons BNN processing.}
    \vspace{-0.4cm}
\label{fig:algo}
\end{figure}

\begin{table} [!h]
\small
    \centering
    \begin{tabular}{c|c|c|c|c|c|c|c|c}
Activations (bits) & 16  & 32 & 64 & 128 & 256 & 512 & 1024 & 2048 \\
\hline
Parallel neur. (max)    & 128 & 64 & 32 & 16  & 8   & 4   & 2    & 1    \\
\hline
Elements number      & 12  & 14 & 16 & 18  & 20  & 22  & 24   & 25  
    \end{tabular}
    % \vspace{-0.3cm}
    \caption{Maximum number of parallel neurons and required number of elements for different activations vector sizes.}
    \vspace{-0.5cm}
    \label{tab:sizes}
\end{table}

N2Net implements the forward pass of a BNN, and assumes that the BNN activations are encoded in a portion of the packet header. The header is parsed as soon as a packet is received, and the parsed activations vector is placed in a PHV's field. Fig.~\ref{fig:algo} summarizes the operations performed by N2Net to implement a 3 neurons BNN. 
The entire process comprises five steps: 
\begin{itemize}[leftmargin=*,nolistsep,noitemsep,topsep=0pt,parsep=0pt,partopsep=0pt]
    \item \textbf{Replication}: in the first step the activations are replicated in the destination PHV as many times as the number of neurons that should be processed in parallel;
    \item \textbf{XNOR and Duplication}: in the second step such fields are processed by the element's RISC processors. The applied actions perform XNOR operations on the activations taking as parameters the neurons' weights. The results are stored twice in the destination PHV. This duplication is required by the implementation of the POPCNT as explained next;
    \item \textbf{POPCNT}: RMT does not support a POPCNT primitive operation. A naive implementation using an unrolled \textit{for} cycle that counts over the vector bits may require a potentially big number of elements. Instead, we adapted a well-known algorithm that counts the number of 1 bits performing additions of partial counts in a tree pattern~\cite{HAKMEM}. The advantage of this algorithm is that it can perform some computations in parallel, while using only shifts, bitwise logic and arithmetic sums. 
    N2Net implements such algorithm combining two pipeline's elements. The first element performs shift/bitwise AND in parallel on the two copies of the input vector. Each copy contains the mutually independent leaves of the algorithm's tree structure. The second element performs the SUM on the outcome of the previous operations. 
    Depending on the length of the activation vector, there may be one or more groups of these two elements, in which case the sum's result is again duplicated in two destination PHV's fields. 
    \item \textbf{SIGN}: the fourth step implements the sign operation verifying that the POPCNT result is bigger or equal to half the length of the activations vector. The result is a single bit stored in the least significant bit of the destination PHV's field.
    \item \textbf{Folding}: the last step folds together the bits resulting from the SIGN operations to build the final Y vector, which can be used as input for a next sequence of 5 steps.
\end{itemize}

\noindent N2Net currently implements fully-connected BNNs taking a model description (number of layers, neurons per layer) and creating a P4 description that modifies/replicates the above five steps as needed. BNN are relatively small models whose weights fit in the pipeline element's SRAMs, however, we are required to pre-configure the weights. This is similar to the BrainWave's approach~\cite{brainwave}.

\noindent \textbf{Evaluation} Our implementation is subject to two main constraints. 
First, the PHV is 512B long. Since we use the PHV to store the BNN input, the maximum activation vector length is 2048 (i.e., half the PHV's size, 256B, since we perform the duplication step). Smaller activation vectors enable the parallel execution of multiple neurons, using the replication step of Fig.~\ref{fig:algo}. For example, with a 32b activation vector, up to 64 neurons can be processed in parallel.
% \footnote{Since the PHV holds data required also to finally forward the packet, these numbers represent upper bounds for an RMT architecture.}. 
Second, the RMT pipeline has 32 elements, and each element can only perform one operation on each of the PHV's fields (for a maximum of 224 parallel operations on independent fields in each element). While we implement the POPCNT leveraging parallelism in order to minimize the number of required elements, we still need $3 + 2log_2(N)$ elements to implement a single neuron, where N is the size of the activations vector (cf. Table\ref{tab:sizes}). 
Using the previous examples, the execution of a neuron with 2048 activations would require 25 elements, while with a 32b activations vector we would take just 13 elements. Furthermore, in this last case the addition of the replication step (i.e., an additional element) would correspond to the parallel execution of up to 64 neurons using only 14 out of the 32 pipeline's elements.

In terms of throughput, an RMT pipeline can process 960 million packets per second. Since we encode in one packet our activations, N2Net enables the processing of 960 million neurons per second, when using 2048b activations. Processing smaller activations enables higher throughput because of parallel processing. 

To put this in perspective, considering the above constraints, we could run about a billion small BNNs per second, i.e., one per each received packet. For instance, we could run 960 million two-layers-BNNs per second, using 32b activations (e.g., the destination IP address of the packet), and two layers of 64 and 32 neurons.
% Still, we have to address a few issues to execute a BNN using a switching chip. First, while bitwise logic is readily available, the computation of the Hamming weight, also called a population count operation, is not provided by the chip's RISC processors. Second, the chip's spatial architecture is designed to process packet headers moving them through the chip's pipeline, with each stage performing just one single operation, although on the entire PHV. Finally, since the pipeline's elements do not share any memory area, it is not possible to share intermediate computation results among the elements. 

% To address the above issues, we devised a general method that matches the chip's parallelism with the intrinsic parallelism of the BNNs computations. 

\section{Challenges and outlook}
N2Net shows that BNN models can be implemented with already available switching chip technology. However, models complexity is limited by relevant constraints.
In turn, this limits the possible applications. We argue these constraints are the outcome of an architecture that was not designed to run NN models, and that supporting them would require relatively cheap design changes.

For example, implementing a simple POPCNT primitive on 32b operands requires few additional logic gates~\cite{domino} but could cut the number of required pipeline's elements. I.e., this would change the 12-25 elements range of Table\ref{tab:sizes} to a 5-10 range. Also, this removes the need for the duplication step, immediately doubling the available space in the PHV, hence doubling the neurons executed in parallel.

Furthermore, the circuitry dedicated to computation (including parsers) accounts for less than 10\% of the switching chip's area. Using 5-10 pipeline's elements to implement BNN computations takes less than a third of that circuitry. Thus, adding a dedicated circuitry for the execution of BNN computations is likely to account for less than a 3-5\% increase in the overall chip area costs.

Overall, we believe N2Net has the potential to open a new interesting field of applications, contributing a novel building block for future networked systems~\cite{daiet}.

% As such, they provide a formidable source of data. For example, it has been shown that the network traffic can be analyzed to discover the presence of security threats, such as malwares~\cite{}, or to build users' profiles~\cite{}.
% At the same time, network devices, such as switches and routers, face great engineering challenges to cope with the exponential growth of network traffic bandwidths~\cite{}. On the one side, they need to move terabits per second of data, processing billions of network \textit{packets} per second. On the other side, network devices have to hold state, usually in the form of forwarding tables, which is read/updated billions of times per second, to ultimately tell how a network packet should be forwarded.   
% In effect, a network device performs a classification task in first place, and then uses the classification result to select a \textit{forwarding action}.

\bibliographystyle{abbrv}
\bibliography{biblio}

\begin{thebibliography}{10}

\bibitem{HAKMEM}
M.~Beeler, R.~W. Gosper, and R.~Schroeppel.
\newblock Hakmem.
\newblock Technical report, Cambridge, MA, USA, 1972.

\bibitem{p4}
P.~Bosshart, D.~Daly, G.~Gibb, M.~Izzard, N.~McKeown, J.~Rexford,
  C.~Schlesinger, D.~Talayco, A.~Vahdat, G.~Varghese, et~al.
\newblock P4: Programming protocol-independent packet processors.
\newblock {\em ACM SIGCOMM CCR}, 44(3):87--95, 2014.

\bibitem{forwardingMetamorphosis}
P.~Bosshart, G.~Gibb, H.-S. Kim, G.~Varghese, N.~McKeown, M.~Izzard, F.~Mujica,
  and M.~Horowitz.
\newblock Forwarding metamorphosis: Fast programmable match-action processing
  in hardware for sdn.
\newblock In {\em Proceedings of the ACM SIGCOMM 2013 Conference on SIGCOMM},
  SIGCOMM '13, pages 99--110, New York, NY, USA, 2013. ACM.

\bibitem{trainingBNN}
M.~Courbariaux and Y.~Bengio.
\newblock Binarynet: Training deep neural networks with weights and activations
  constrained to +1 or -1.
\newblock {\em CoRR}, abs/1602.02830, 2016.

\bibitem{SwitchPaxos}
H.~T. Dang, M.~Canini, F.~Pedone, and R.~Soul{\'e}.
\newblock Paxos made switch-y.
\newblock {\em SIGCOMM Comput. Commun. Rev.}, 46(2):18--24, May 2016.

\bibitem{bnn}
I.~Hubara, M.~Courbariaux, D.~Soudry, R.~El-Yaniv, and Y.~Bengio.
\newblock Binarized neural networks.
\newblock In D.~D. Lee, M.~Sugiyama, U.~V. Luxburg, I.~Guyon, and R.~Garnett,
  editors, {\em Advances in Neural Information Processing Systems 29}, pages
  4107--4115. Curran Associates, Inc., 2016.

\bibitem{NetCache}
X.~Jin, X.~Li, H.~Zhang, R.~Soul{\'e}, J.~Lee, N.~Foster, C.~Kim, and
  I.~Stoica.
\newblock Netcache: Balancing key-value stores with fast in-network caching.
\newblock In {\em Proceedings of the 26th Symposium on Operating Systems
  Principles}, SOSP '17, pages 121--136, New York, NY, USA, 2017. ACM.

\bibitem{bitwiseNN}
M.~Kim and P.~Smaragdis.
\newblock Bitwise neural networks.
\newblock {\em CoRR}, abs/1601.06071, 2016.

\bibitem{learnedIndexes}
T.~Kraska, A.~Beutel, E.~H. Chi, J.~Dean, and N.~Polyzotis.
\newblock The case for learned index structures.
\newblock 2017.

\bibitem{switchKV}
X.~Li, R.~Sethi, M.~Kaminsky, D.~G. Andersen, and M.~J. Freedman.
\newblock Be fast, cheap and in control with switchkv.
\newblock In {\em 13th {USENIX} Symposium on Networked Systems Design and
  Implementation ({NSDI} 16)}, pages 31--44, Santa Clara, CA, 2016. {USENIX}
  Association.

\bibitem{silkroad}
R.~Miao, H.~Zeng, C.~Kim, J.~Lee, and M.~Yu.
\newblock Silkroad: Making stateful layer-4 load balancing fast and cheap using
  switching asics.
\newblock In {\em Proceedings of the Conference of the ACM Special Interest
  Group on Data Communication}, SIGCOMM '17, pages 15--28, New York, NY, USA,
  2017. ACM.

\bibitem{brainwave}
Microsoft.
\newblock Microsoft unveils project brainwave for real-time ai.
\newblock
  \url{https://www.microsoft.com/en-us/research/blog/microsoft-unveils-project-brainwave/}.

\bibitem{XNOR-net}
M.~Rastegari, V.~Ordonez, J.~Redmon, and A.~Farhadi.
\newblock Xnor-net: Imagenet classification using binary convolutional neural
  networks.
\newblock {\em CoRR}, abs/1603.05279, 2016.

\bibitem{daiet}
A.~Sapio, I.~Abdelaziz, A.~Aldilaijan, M.~Canini, and P.~Kalnis.
\newblock In-network computation is a dumb idea whose time has come.
\newblock In {\em Proceedings of the 16th {ACM} Workshop on Hot Topics in
  Networks, Palo Alto, CA, USA, HotNets 2017, November 30 - December 01, 2017},
  pages 150--156, 2017.

\bibitem{flexswitches}
N.~K. Sharma, A.~Kaufmann, T.~Anderson, A.~Krishnamurthy, J.~Nelson, and
  S.~Peter.
\newblock Evaluating the power of flexible packet processing for network
  resource allocation.
\newblock In {\em 14th {USENIX} Symposium on Networked Systems Design and
  Implementation ({NSDI} 17)}, pages 67--82, Boston, MA, 2017. {USENIX}
  Association.

\bibitem{domino}
A.~Sivaraman, A.~Cheung, M.~Budiu, C.~Kim, M.~Alizadeh, H.~Balakrishnan,
  G.~Varghese, N.~McKeown, and S.~Licking.
\newblock Packet transactions: High-level programming for line-rate switches.
\newblock In {\em ACM SIGCOMM '16}, ACM SIGCOMM '16, pages 15--28. ACM, 2016.

\end{thebibliography}

\end{document}